# Multi-Channel Data Acquisition System with Absolute Time Synchronization

*Przemysław Włodarczyk[a], Szymon Pustelny[b,c], Dmitry Budker[c], and Marcin Lipiński[a]*

[a] Department of Electronics, AGH University of Science and Technology, Mickiewicza 30, 30-059 Krakow, Poland
[b] Institute of Physics, Jagiellonian University, Reymonta 4, 30-059 Krakow, Poland
[c] Department of Physics, University of California at Berkeley, Berkeley, California 94720-7300, USA

**Abstract**

A low-cost, stand-alone Global-Positioning-System-time-synchronized data acquisition system is described. The constructed prototype allows recoding up to four analog signals with a 16-bit resolution in variable ranges and a maximum sampling rate of 1000 S/s. Additionally, two digital readouts of external sensors can be acquired. A complete data set is stored on a Secure Digital (SD) card or transmitted to a computer using Universal Serial Bus (USB). The estimated time accuracy of the data acquisition is better than ±0.2 μs. The device is envisioned for the use in a global distributed sensor network (the Global Network of Optical Magnetometers for Exotic physics – GNOME), whose aim is to search for new particles and interactions.

## 1. Introduction

Many modern experiments require measurements of various physical quantities in distant locations. An example of such measurements is geophysical studies, in particular, investigations of Earth's seismic activity. Synchronous detection of seismic waves in different geo-observatories not only provides information about earthquake magnitude and epicenter but also enables imaging of deep geological structures (seismic tomography) [1]. Time resolution of such measurements, however, can be relatively low (~$10^{-4}$ s) due to low speed of the seismic waves (~10 km/s) [2]. Various astronomical and astrophysical observations require much better time resolution (as the propagation speed of the signals is typically comparable to the speed of light). For example, a microsecond precision is needed to enhance performance of gravitational-wave (see, for example, Ref. [3]) and neutrino detectors [4], but the most sophisticated astronomical studies require measurements with picosecond time resolution [5].

Nowadays, on-site measurements with time resolution at a nanosecond level are routinely performed. Such precision may be obtained by application of commercial atomic clocks [6]. Moreover, sub-nanosecond resolution may be achieved using ultra-cold atoms (atomic fountains [7] and optical lattices [8]), trapped ions [9], or active maser standards [10]. The most precise clock ever demonstrated is based on $Al^+$ ions and is being operated at the National Institute of Standard and Technology in Boulder, Colorado. The clock provides accuracy of ~$10^{-17}$ [9].

While accurate time reference may be provided in various ways, distribution of this reference and high-accuracy synchronization of distant experiments is a challenge. Currently, several techniques of time distribution are used. For example, time servers distribute timestamps over the Internet [11], which facilitates data exchange and computer-network communication. Global accuracy of this scheme, however, is not high (a few tens of milliseconds). Better precision can be achieved by radio-clock synchronization [12]. Various radio stations (WWV Colorado, RWM Moscow, BPM Pucheng, etc.) broadcast time sequences that may be used for that purpose. Unfortunately, none of the radio-clock signals can be received everywhere on Earth, thus synchronization of experiments separated by thousands of kilometers must exploit signals delivered by different stations and needs to be based on an assumption of mutual synchronization of the stations' clocks. Beside the limited precision of such synchronization, additional uncertainties arise from fluctuations of propagation conditions of the radio waves through the atmosphere. As a result, the accuracy of radio synchronization can be realized at a millisecond level (after accounting for propagation delays). The most precise means of time distribution exploits fiber-optic network (see Ref. [13] and references therein). Using this approach, time may be distributed with the accuracy better than a picosecond [14]. This solution, however, requires the use of fiber networks, which are not always available (e.g., for in-field measurements) and application of dedicated equipment (transducers, regenerators, amplifiers, etc.), which is not widely available. This limits the universality of the solution. Somewhat worse precision may be provided by global positioning system (GPS). Each GPS satellite is equipped with four atomic clocks that are being referenced to terrestrial master clock. The problem with distribution of time is that it is affected by varying transmission properties of Earth's atmosphere. This can be alleviated by detection of the GPS signals from many satellites and application of special algorithms enabling time precision at a nanosecond level. An important advantage of this solution is global 'visibility' of GPS satellites allowing delivery of timestamp all over the world.

In this paper, we describe an inexpensive autonomous time-stamped data acquisition system. The system enables simultaneous measurements of four analog and two digital channels with time provided by the GPS. The signals may be recorded with a sampling rate up to 1000 samples per second (1 kS/s), and they can be stored in a built-in memory card. The internal memory enables fully computer-independent operation of the system; the device is controlled and operated using a set of buttons and a screen mounted at the front panel of the device. Such operation, however, is only possible for a finite time due to the limited capacity of the memory card. Due to this reason, the device can be also operated as a part of computer system. In that case, Universal Serial Bus (USB) allows for device-computer communication, in particular, setting up acquisition parameters, and enables efficient data transfer between two devices. During such operation, the memory card works as a data storage buffer, i.e., ensures continuous data acquisition in case of lost communication with the computer.

**2. Data acquisition system**

A block diagram of our data acquisition system is shown in Fig. 1. The system is based on ARM7 core Atmel microcontroller, whose role consists in handling time reference, controlling and synchronizing data acquisition, storing data at the memory card, and communication with a computer. The microcontroller additionally handles LCD screen and monitors the control buttons mounted at the front panel of the device.

In the system, the time reference is provided by a GPS time receiver (Trimble Resolution T). The module generates the pulse marking the beginning of a second (pulse per second, PPS). To obtain precise timing of second's beginning, the position of the GPS antenna needs to be precisely known. This information is obtained automatically prior to time synchronized data acquisition by averaging positions from succeeding measurements (2000 position measurements). After acquiring information about the location of the device, the reference pulse is synchronized to GPS or UTC time with an accuracy of 45 ns ($3\sigma$) [15].

The GPS-module time-reference pulses are transmitted to the microprocessor. Each pulse is followed by a serial-port message containing exact time information. Additionally, the message contains information that may be used to estimate time reliability, e.g., the number of 'visible' satellites and warnings reporting any problems. Finally, it contains information about position (longitude, latitude, and altitude), temperature, and other auxiliary data regarding the module.

The acquisition system has four isolated analog input channels. Each channel is equipped with a precision selectable-gain preamplifier designed with OPA277 operational amplifier and a 16-bit analog-to-digital converter ADS8507 (ADC). Application of the preamplifiers ensures versatility of the device, i.e., an entire measurement range of the converters is used. The microcontroller, clocked with its local oscillator at a frequency of 48 MHz, generates pulses that trigger the ADCs. The triggering pulses are synchronized with the GPS reference pulse to reduce slow drifts of the internal clock. After each trigger, when ADCs are ready, converted data are read with Serial Peripheral Interface Bus (SPI) in a daisy-chain scheme [16]. The channels are read with a sampling rate of 1 kS/s and in case of a lower sampling rate the unnecessary samples are not saved. Simultaneously, slower digital sensors (e.g., thermometers, magnetometers, etc.) are read with a maximum sampling rate of 50 samples per second (50 S/s).

The collected data are saved on the Secure Digital (SD) card formatted with the FAT32 file system, which acts as a buffer. The data is also sent to the computer (if available) via USB. Support of the SD card, FAT32, and USB is realized based on open-source libraries ([17, 18]).

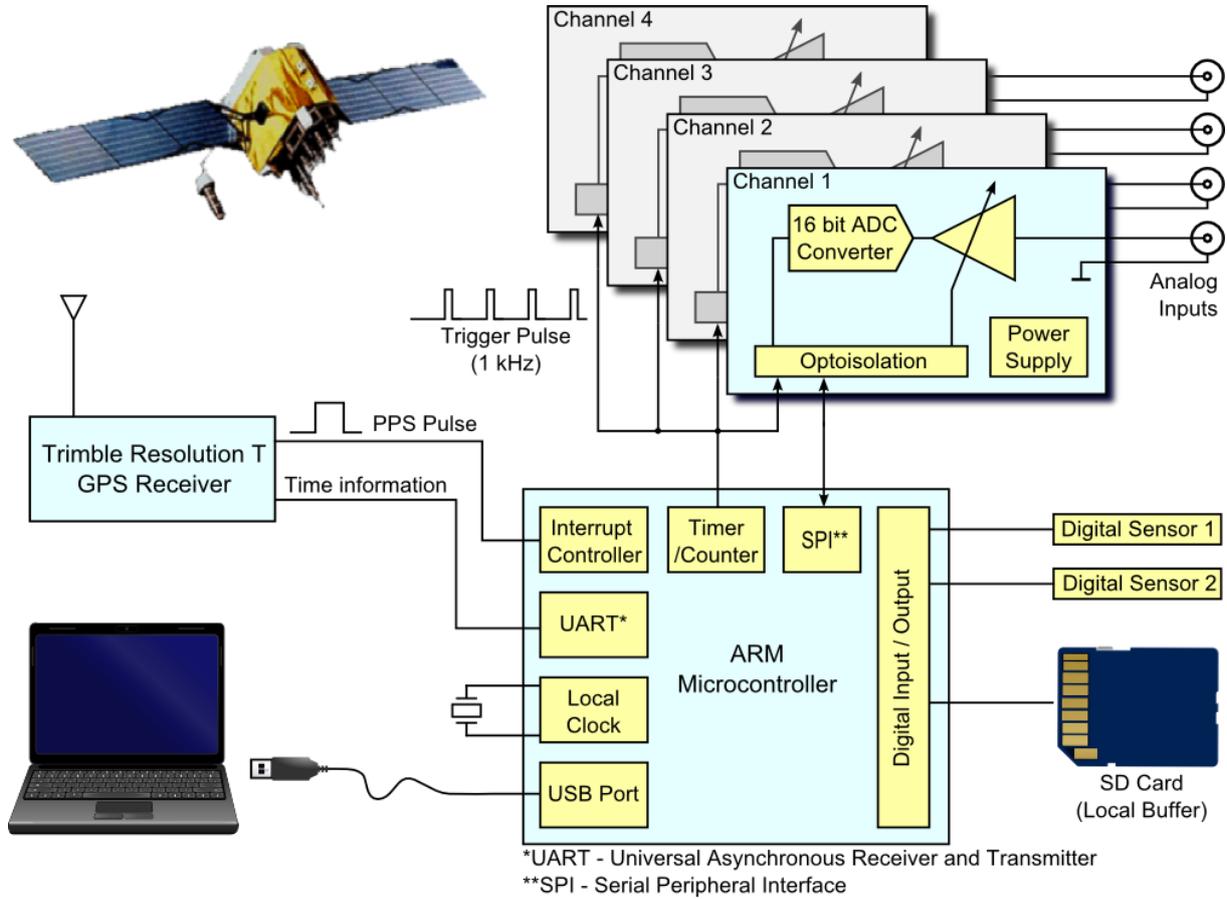

**Fig. 1.** Block diagram of the data acquisition system. The GPS receiver provides synchronization pulses every second. Microcontroller generates 1 kHz pulses used to trigger ADCs, which samples input signals at up to 4 channels. Sampled data preceded by a header containing valid GPS information are buffered on an SD memory card and are transited to a computer via USB.

**3. Synchronization algorithm**

Figure 2 shows the time sequence used in our data acquisition system. An acquisition is initiated by the rising edge of a time-reference pulse (PPS). The pulse triggers the highest-priority interrupt, called FIQ [19], which starts a service routine. The routine resets the timer/counter used for triggering the ADCs. Within the next second, the timer/counter is reset every time it reaches $C$ clock cycles. The value of $C$ is chosen in such a way that $C$ clock cycles correspond to 1 ms (ideally 48 000 cycles, see below). When the timer/counter is reset, ADCs' trigger pulse is generated and the sampling and conversion is initiated. It should be noted that in our system, the signals in all channels are sampled for only 5 ns, while their conversion takes up to 25 μs. After the conversion, the microcontroller reads the converted data, which is realized in a high-priority interrupt.

Measurement of the last (1000$^{th}$) sample within a given second disables the timer/counter reset (it still increases every clock cycle) and introduces the microcontroller into the awaiting mode (waiting for the next synchronization pulse). When the pulse arrives, the timer/counter value is first read and then it is immediately reset. If the local clock has its nominal frequency (48 MHz), the read value should be equal to $C$, which corresponds to 1 ms. If it is not, the difference between $C$ and the number of measured cycles is used to adjust $C$. This technique compensates for possible drifts of the local-oscillator frequency, i.e., it ensures that recorded samples are uniformly distributed over time with 1 ms spacing between them.

In addition to the detection of the analog signals, after a fixed number of ADC samples, data from auxiliary digital sensors are read. This transmission has a lower priority and may be interrupted by the ADC service routine. The update rate of the digital sensors depends on their speed but is typically low, e.g., in our system temperature is sampled once a second and magnetic field is read with 50 S/s sampling rate.

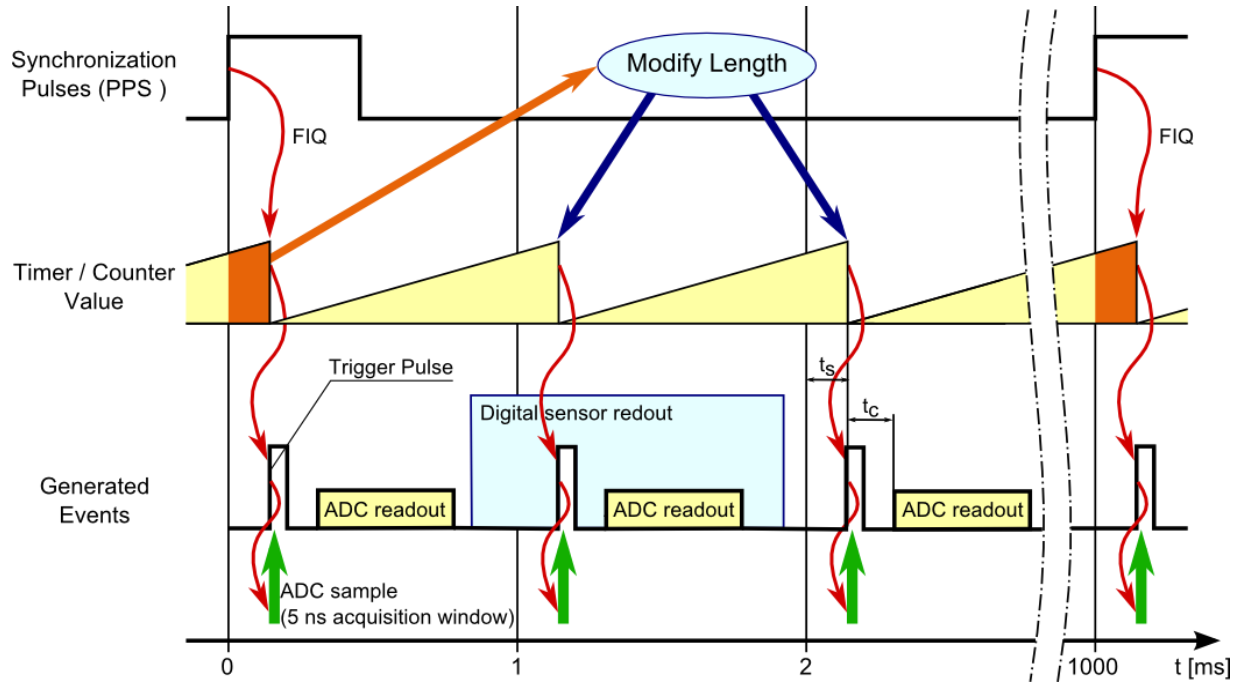

**Fig. 2.** Time sequence of GPS synchronized data acquisition (not to scale). The FIQ interrupt zeros the timer/counter and synchronizes the local clock to the GPS time reference. The timer/counter triggers the ADCs every millisecond (in practice every $C$ clock cycles). ADCs' trigger pulse initiating analog data acquisition has a software delay $t_s$ due to entering interrupt. The signal is sampled for about 5 ns (vertical arrow pointing up) and then converted by the ADCs for the conversion time $t_c$. Next, the data are shifted out of the ADCs. When all ADC data are read, lower-priority digital sensor may be serviced (big rectangle). To correct for the local-clock frequency drift, the timer/counter value is measured at the end of each second. If the counter readout differs from the expected value $C$ (shaded part of the triangle), the clock division of the second (aimed to be 1 ms periods) is adjusted to account for the drift. It ensures more uniform distribution of the ADC samples.

During the entire second, data from the analog inputs and digital sensors are stored in a buffer located in microcontroller's random-access memory (RAM). After the second acquisition is completed, the collected samples are combined with GPS data (time, position, warnings, etc.) and saved at the Secure Digital (SD) card, which acts as a semi-permanent buffer. In our scheme, the data are saved as minute-long files (file size up to 1.8 MB) in a human readable text format. The files saved on the SD card are, starting with the oldest, continuously sent to the computer via USB. If USB is not connected and the SD card space is low, the oldest files are overwritten (for a 4 GB SD card, the buffer is up to 20 hours long). After reestablishing the USB transmission, the data exchange is resumed. The USB data transfer is realized as low priority process running continuously in the time between ADC and digital sensor readouts.

**4. Precision of GPS synchronized data acquisition**

Although data acquisition is synchronized with the GPS time reference, there are various contributions that introduce delays[1] and/or uncertainties in our system (Table 1).

One source of the delay in the system is finite time of propagation of the GPS signal in the antenna cable (4 ns/m). It should be noted, however, that to a large extent, this delay is constant in a given configuration so it is not a source of significant uncertainty. Moreover, in our system, the direct communication with Trimble Resolution T in the, so-called, service mode, ensures that the delay in the cable can be compensated. Nonetheless, the delay uncertainty, which exists due to the drift of cable parameters (e.g., with humidity or temperature) are harder to compensate. We estimate that in a worst-case scenario, the uncertainty should not exceed 10% of the total cable delay.

---

[1] Note that the delays are constant, i.e., do not fluctuate over time.

The next source of uncertainty is introduced by the time of arrival of the GPS synchronization pulse; fluctuations of the atmosphere's properties change the propagation conditions and hence speed of the electromagnetic wave in the atmosphere. According to the manual of the GPS module [15], the 3σ time uncertainty is 45 ns.

The largest error introduced by the software arises from the finite time $t_s$ that is needed to call the synchronization routine, which typically takes ~1.5 μs. Due to this initial delay, all the trigger pulses in the following second are delayed. It should be noted that the delay $t_s$ is spread from one second to another because the time needed to enter an interrupt depends on the instantaneous processor state. The total software delay is between 1450 ns and 1630 ns, which corresponds to an uncertainty of 180 ns. This uncertainty can be eliminated by modifying the system so that the trigger pulse provides hardware reset of the timer/counter.

Onboard propagation delays are estimated on the basis of catalog datasheets of electronic components that are used in the project. The main contribution to this error is the optocoupler (6N137) propagation delay. There is also a CMOS buffer used in the signal path, which adds typically 20 ns.

Another source of delay originates from analog-to-digital conversion. Each ADC measurement is delayed relative to the trigger pulse by the time called aperture delay, which is typically 40 ns. Its jitter is 20 ps, which is negligible in the described system.

The error budget shows that the samples are delayed by about 1.8 μs relative to the GPS time and the maximum time uncertainty is better than ±0.2 μs. It should be stressed that this is still almost three orders of magnitude smaller than the maximum sampling rate of the system. Moreover, the main delay occurring in the prototype can be eliminated making acquisition with time accuracy better than ±0.2 μs feasible.

Table 1. Synchronization error budget.

| Source of Error | Min. Delay | Max. Delay | Max. Uncertainty |
|---|---|---|---|
| Antenna cable delay* | 0 | + 200 ns (50 m of cable) | ±10 ns |
| GPS synchronization error (3σ) | -45 ns | 45 ns | ±45 ns |
| Program delays ($t_s$) ** | 1450 ns | 1630 ns | ±90 ns |
| Estimated onboard propagation delays | 40 ns | 110 ns | ±35 ns |
| ADC Aperture delay | 40 ns | 40 ns | ±20 ps |
| **Total** | **1485 ns** | **2025 ns** | **±180 ns** |

* The system provides compensation if the cable delay is known.
** Possible to eliminate in the next version, hardware changes required.

## 5. Real data acquisition

To verify the operation of the device, we recorded the time signal transmitted by the National Institute of Standards and Technology (NIST) radio station located at Fort Collins, Colorado (WWV Colorado). The WWV signal is quite complex. It contains, among others, 1 kHz bursts. The first burst of each minute is 800 ms long and heard as a 'beep'. The other second pulses are much shorter (5 ms) and heard as 'ticks'. At the end of every minute, a voice announcement is made giving the time of the following minute [20].

To obtain the minute and second pulses, the WWV signal, received with a battery powered receiver, was filtered with a 1 kHz bandpass filter and rectified. After this initial processing, the signal was recorded using our system.

Figure 3 shows the recorded WWV signals. Since the signal shows a beginning of a minute, thus a first 800 ms minute pulse followed by 5 ms second pulses are recorded [Fig. 3a]. The second ticks are doubled as the difference between the Coordinated Universal Time and the Universal Time is coded that way [20]. When zoomed [Fig. 3b], it can be seen that the recorded minute pulse is delayed by 6 ms. The distance between the NIST transmitter and Berkeley, where the signal was recorded is 1500 km, but radio waves travel over a longer distance due to ionosphere reflections. The propagation delay, computed by Janusz Młynarczyk using IRI-2007 ionosphere model [21], was 5.4 ±0.3 ms. An additional 420 μs delay is introduced by radio receiver and filter used to separate time pulses, which was confirmed by measurement with a simulated WWV signal.

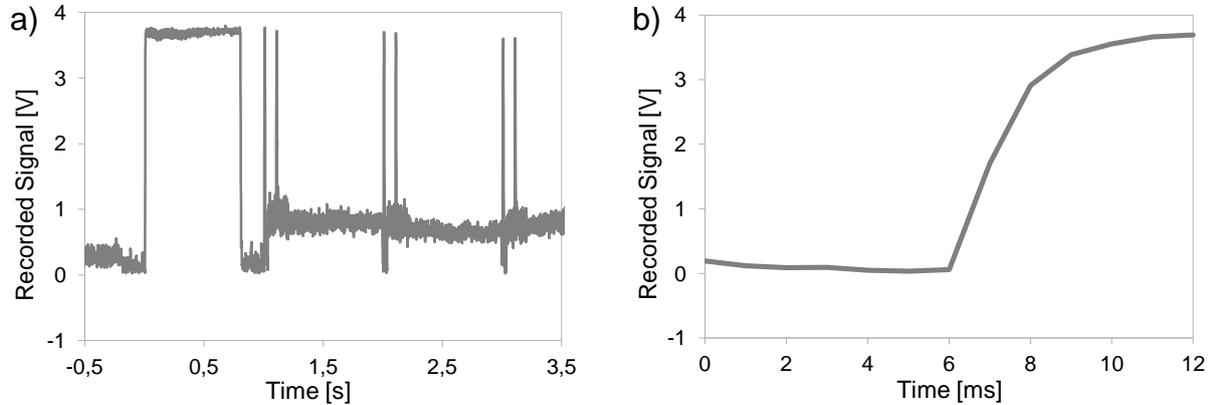

**Fig. 3.** NIST time signal recorded with our system. The system operates at a sampling rate of 1 kHz and input range of ±5 V. Abscissa 0 refers to 2012.12.04 04:33:00 UTC. The minute synchronization pulse is delayed due to the radio-wave propagation time (6 ms).

**6. Summary**

We described a low-cost, stand-alone GPS-time-synchronized data acquisition system. The constructed prototype (Fig. 4) allows recoding up to four analog signals with a 16-bit resolution and a maximum sampling rate of 1000 S/s. All channels have variable gain resulting in ±1.25 V to ±10 V input range. Additionally, two digital readouts of external sensors can be measured. A complete data set can be stored on a SD card and transmitted to a computer using USB. The estimated time accuracy of the data acquisition with respect to GPS or UTC time is better than ±0.2 μs with ~1.5 μs constant delay. Moreover, improvements of the accuracy and increase of the sampling rate may be achieved in the future by upgrades of the system.

The described device is envisioned to be used in the Global Network of Optical Magnetometers for Exotic physics (GNOME) being developed for a purpose of searching for new particles and interactions [22]. The network will eventually include several optical magnetometers (see, for example, [23]), whose signals will be recorded simultaneously. The recorded data will be searched for the correlated signals corresponding to the events with global impact. The time difference between occurrences of the same event in various locations will be used to compute the direction and the speed of the detected object [24].

**Acknowledgments**

The authors acknowledge Janusz Młynarczyk of the AGH University of Science and Technology, Poland, for his help in calculating propagation of electromagnetic wave through the atmosphere. This research has been supported by the National Science Foundation Grant No. PHY-1068875 and the grant of the National Centre for Research and Development within the Leader Programme.

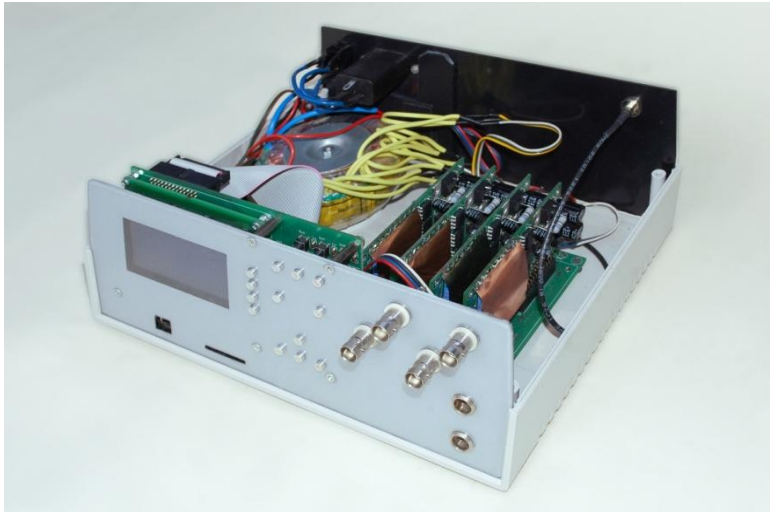

**Fig. 4.** Photo of the prototype of the data acquisition system.